\documentclass[10pt,a4paper]{article}
\usepackage{bm, amssymb,pifont,cancel, amsmath,comment,color,authblk}
\usepackage[dvips]{graphicx}
\usepackage{multicol}
\newcommand{\Da}{{\cal D}_{\alpha}}
\newcommand{\Dad}{\bar{\cal D}_{\dot{\alpha}}}
\newcommand{\uDa}{{\cal D}^{\alpha}}
\newcommand{\uDad}{\bar{\cal D}^{\dot{\alpha}}}
\begin{document}
\begin{flushright}
WU-HEP-14-09
\end{flushright}
\begin{center}
{\Large{\bf{Inflation in supergravity without K\"ahler potential}}}\\ 
\vskip 6pt
\large{Shuntaro Aoki}{\renewcommand{\thefootnote}{\fnsymbol{footnote}}\footnote[1]{E-mail address: shun-soccer@akane.waseda.jp}}, and \large{Yusuke Yamada}{\renewcommand{\thefootnote}{\fnsymbol{footnote}}\footnote[2]{E-mail address: yuusuke-yamada@asagi.waseda.jp}}\\
\vskip 4pt
{\small{\it Department of Physics, Waseda University,}}\\
{\small{\it Tokyo 169-8555, Japan}}
\end{center}
\begin{abstract}
We propose a new class of inflation models in supergravity with higher derivative terms. In those models, the K\"ahler potential does not contain the inflaton multiplet, but a supersymmetric derivative term does. In the models, inflation is effectively driven by a single scalar field with a standard kinetic term and a scalar potential. Remarkably, the so-called $\eta$ problem does not exist in our models.
\end{abstract}
\begin{multicols}{2}
\section{Introduction}
The inflationary paradigm~\cite{Guth:1980zm,Starobinsky:1980te} is a part of the standard cosmology. Especially, in slow-roll inflation models~\cite{Linde:1983gd}, the primordial curvature perturbations are naturally produced by the quantum fluctuation of (at least) one scalar field called inflaton during inflation. The spectrum of scalar perturbation predicted by a slow-roll model is almost flat, which is compatible with the cosmic microwave background (CMB) observation results.

Inflation models in supergravity (SUGRA) have been studied through many works. It is interesting that SUGRA is an effective theory of superstring, which is a possible candidate for a unified theory including gravity. 

In this work, we propose a new class of inflation models in SUGRA including higher derivative interactions. Higher derivative terms always arise from e.g. the Dirac-Born-Infeld action describing D-brane dynamics in string theory. Some models with higher derivative terms have been constructed in SUGRA~\cite{Koehn:2012ar,Farakos:2012je}. Higher derivative terms in SUGRA can be described by using the supersymmetry~(SUSY) covariant derivative operator of SUSY multiplets. We call a SUGRA term including SUSY derivative operators a SUSY derivative term in this letter.  The components of SUSY derivative terms contain the higher derivative interactions. 

In the standard SUGRA action without SUSY derivative terms, dynamical chiral multiplets should be contained in a K\"ahler potential. However, as we will show, the kinetic terms of chiral multiplets also arise from SUSY derivative terms.

 We will discuss inflation models in which the inflaton multiplet is not included in the K\"ahler potential. Surprisingly, the so-called $\eta$ problem, which spoils the flatness of the inflation potential, does not exist in our models. It is also remarkable that the effects of the higher derivative terms of the inflaton become negligibly small on the slow-roll trajectory. Consequently, the inflation is effectively driven by a single scalar field with a standard kinetic term and a scalar potential.  We will show two inflation models, the chaotic inflation\cite{Linde:1983gd} and the Starobinsky inflation~\cite{Starobinsky:1980te}. Especially, the chaotic inflation in our model predicts a characteristic spectrum of the perturbation, which can be tested by forthcoming observations.

The remaining parts of this letter are consisted as follows. First, in Sec.~\ref{setup}, we will show the action we will discuss. After that, choosing the two types of superpotential, we will construct the chaotic and the Starobinsky like inflation models, and show the cosmological parameters in those models in Sec.~\ref{chaotic} and \ref{Starobinsky}. Finally we will conclude in Sec.~\ref{con}. In Appendix.~\ref{A}, the cosmological parameters in our models are discussed.
\section{Inflation with higher derivative terms}\label{model}
\subsection{Setup}\label{setup}
To clarify the SUGRA system discussed below, we first show the corresponding action in superspace. It is described by 
\begin{align}
S=&\int d^4\theta K(\hat{S},\bar{\hat{S}})+\left(\int d^2\theta W(\hat{\Phi},\hat{S})+{\rm h.c.}\right)\nonumber\\
&+\int d^4\theta C D_\alpha\hat{\Phi}  D^{\alpha}\hat{S}\bar{D}_{\dot{\alpha}}\bar{\hat{\Phi}}\bar{D}^{\dot{\alpha}}\bar{\hat{S}},\label{gSUSY}
\end{align}
where $\hat{S}$ and $\hat{\Phi}$ are chiral multiplets, $K$ and $W$ are K\"ahler and super-potential respectively, $C$ is a constant, $D_\alpha$ and $\bar{D}_{\dot{\alpha}}$ denote SUSY covariant derivatives, and $\alpha,\dot{\alpha}$ are spinor indices. Here and the following, we use the unit~$M_{P}=1$ where $M_P=2.4\times10^{18}$GeV is the reduced Planck mass. We will denote the scalar component of each multiplet by the character without a hat below. 

The last term in Eq.~(\ref{gSUSY}) is a SUSY derivative term. It leads ghost-free interactions with respect to the bosonic sector discussed in Ref.~\cite{Koehn:2012ar}. In the well-known SUGRA models without SUSY derivative terms, a dynamical chiral multiplet should be contained in a K\"ahler potential $K$ because superpotential terms does not give kinetic terms for chiral multiplets. However, in our case, the SUSY derivative term in Eq.~(\ref{gSUSY}) gives $S$ and $\Phi$ not only higher derivative terms, but also standard kinetic terms in non-canonical forms as we will show. 

Here, we choose the following $K$ and $W$~\cite{Kawasaki:2000yn,Kallosh:2010xz}, 
\begin{eqnarray}
K&=&|\hat{S}|^2,\label{K}\\
W&=&f(\hat{\Phi}) \hat{S},
\end{eqnarray}
where $f(\hat{\Phi})$ is a holomorphic function of $\hat{\Phi}$. We emphasize that the K\"ahler potential~(\ref{K}) does not contain $\hat{\Phi}$, which we identify as the inflaton multiplet. $\hat{S}$ becomes the Goldstino multiplet during inflation. With a minimal K\"ahler potential $K=|\hat{S}|^2$, $S$ often has its light or tachyonic mass during inflation, and therefore the quantum fluctuation of $S$ may become a source of the curvature perturbation as discussed in Ref.~\cite{Kawasaki:2000yn,Demozzi:2010aj}. Recently, the authors of Ref.~\cite{Antoniadis:2014oya} showed the absence of such a situation in the case that $\hat{S}$ is the Volkov and Akulov supermultiplet~\cite{Volkov:1973ix}, which satisfies $\hat{S}^2=0$.\footnote{We can construct a similar model with an unconstrained multiplet instead of $\hat{S}$ if its scalar component has a sufficiently heavy mass. However, in that case, the equations of motions of auxiliary fields are complicated and the solutions are not determined uniquely~\cite{Koehn:2012ar}.} Then, $S$ is identically $0$ because it is the bilinear of the fermionic component of $\hat{S}$~\cite{Antoniadis:2014oya}. 

In conformal SUGRA~\cite{Kugo:1982cu,Kugo:1983mv}, the action corresponding to one in Eq.~(\ref{gSUSY}) can be expressed as follows, 
\begin{align}
S_{\rm SG}=&\frac{1}{2}\left[S_0\bar{S}_0 (-3e^{-\frac{K}{3}})\right]_D\nonumber\\
&+\left[S_0^3W\right]_F+[C\Da \Phi\uDa S\Dad \bar{\Phi}\uDad \bar{S}]_D,\label{Sconf}
\end{align} 
where $[\cdots]_{D,F}$ denote the superconformal D- and F-density formulae~\cite{Kugo:1982cu}, $S_0$ is a chiral compensator, and $\Da$ denotes the superconformal spinor derivative corresponding to $D_\alpha$ in Eq.~(\ref{gSUSY})~\cite{Kugo:1983mv}.\footnote{We note that the last term in Eq.~(\ref{Sconf}) is manifestly K\"ahler invariant. The K\"ahler transformation, $K\rightarrow K+\Lambda+\bar{\Lambda}$, $W\rightarrow We^{\Lambda}$, is the redefinition of the compensator $S_0$ in conformal SUGRA. Obviously, the last term in Eq.~(\ref{Sconf}) is not coupled to the compensator, and therefore it is inert under the K\"ahler transformation. However, the sigma model diffeomorphism, implies that $C$ in Eq.~(\ref{Sconf}) should be the tensor of the target space.} In the following, we only discuss the bosonic action, and then we can put the usual superconformal gauge fixing conditions~\cite{Kugo:1982cu,Kugo:1983mv}, e.g. $S_0=\bar{S}_0=e^{K/6}$ and obtain the SUGRA action in Einstein frame. 

After eliminating the compensator $S_0$ and the auxiliary field $A_\mu$, which is one of the gauge fields of the superconformal symmetry, we obtain the following action, 
\begin{align}
\mathcal{L}=&K_{S\bar{S}}F^S\bar{F}^{\bar{S}}-3F^0\bar{F}^0\nonumber\\
&+\left(W_SF^S+{\rm h.c.}\right)\nonumber\\
&+32CF^\Phi F^S\bar{F}^{\bar{\Phi}}\bar{F}^{\bar{S}}-16C\partial_\mu\Phi \partial ^\mu\bar{\Phi}F^S\bar{F}^{\bar{S}},
\end{align}
where $F^{0,S,\Phi}$ are the auxiliary components of $\hat{S}_0,\hat{S},$ and $\hat{\Phi}$ respectively and we used the condition $K=K_S=W_\Phi=W=0$ because $S\equiv 0$. The solutions for the equations of motion of the auxiliary fields are simply given by 
\begin{align}
&F^0=0,\\
&F^{\Phi}=0,\\
&F^S=-\frac{\bar{W}_{\bar{S}}}{1-16C\partial_\mu\Phi\partial^\mu\bar{\Phi}},
\end{align}
where we have used $K_{S\bar{S}}=1$.

By substituting the on-shell expressions of F-terms into the action, we obtain the effective Lagrangian as follows,
\begin{align}
\mathcal{L}=\frac{V}{A^2}\tilde{X}-\frac{V}{A}\left(2-\frac{1}{A}\right),\label{Leff}
\end{align}
where $V\equiv|W_S|^2=|f(\Phi)|^2$, $\tilde{X}\equiv-16C \partial_\mu\Phi\partial^\mu\bar{\Phi}$ and $A\equiv (1+\tilde{X})$. We can redefine the complex scalar $\Phi$ as follows,
\begin{align}
\varphi\equiv4\sqrt{C}\int f(\Phi)d\Phi.
\end{align}
Then the Lagrangian~(\ref{Leff}) can be rewritten as
\begin{align}
\mathcal{L}=\frac{X}{A^2}-\frac{V}{A}\left(2-\frac{1}{A}\right),
\end{align}
where $X\equiv \partial_\mu \varphi\partial^\mu\bar{\varphi}=V\tilde{X}$. 

Although the action contains the higher order terms of $X$, the cosmological parameters are same with ones in the models without higher derivative action as shown in Appendix.~\ref{A}. Therefore, on the slow-roll trajectory, we can approximate the Lagrangian as follows, 
\begin{align}
\mathcal{L}\sim X-V,
\end{align}
where we used the approximation $A\sim 1$. The approximated action is one with the standard kinetic and the potential terms of $\varphi$, and we will use this action in the following discussion.

It is worth to remark the absence of the $\eta$ problem in this model. In the standard SUGRA models, the F-term scalar potential is given by,
\begin{eqnarray}
V_F=e^K\left(K^{I\bar{J}}D_IWD_{\bar{J}}\bar{W}-3|W|^2\right),
\end{eqnarray} 
where $D_IW=W_I +K_IW$, and $K^{I\bar{J}}$ is the inverse of $K_{I\bar{J}}$. In the case that K\"ahler potential is given by $K=|\Phi|^2+\cdots$, the scalar potential becomes steep due to the factor $e^K$. In our case, however, the inflaton $\Phi$ is not contained in the K\"ahler potential, and therefore the $\eta$ problem does not exist.\footnote{From the conformal SUGRA viewpoint, we can also understand the reason why the $\eta$ problem does not exist in our model. In our case, the last term in Eq.(\ref{Sconf}) does not contains the mixing between Ricci scalar and the inflaton. In this case, the $\eta$ problem does not occur~\cite{Abe:2014opa}. } 
\subsection{Chaotic inflation}\label{chaotic}
We consider the following function $f(\Phi)$,
\begin{align}
f(\Phi)=\lambda_n\Phi^n,
\end{align} 
where $n\geq1$, $\lambda_n$ is a coupling constant. Then the scalar $\varphi$ can be written as
\begin{align}
\varphi=\frac{4}{n+1}\sqrt{C}\lambda_n\Phi^{n+1}.
\end{align}
We identify $\sqrt{2}|\varphi|$ as the inflaton denoted by $\phi$, and we can write down the scalar potential $V$ as follows,
\begin{align}
V=\tilde{\lambda}_n^2\phi^{2n/(n+1)},\label{Vc}
\end{align}
where $\tilde{\lambda}_n^2\equiv \left(2^{-n}(16C)^{-n}(n+1)^{2n}\lambda_n^2\right)^{1/(n+1)}$.  Note that this model resembles to the running kinetic inflation model~\cite{Nakayama:2010kt}, however, the scalar potential is highly restricted in our case.

Surprisingly enough, the effective potential is restricted from the linear potential to the quadratic one, even when the power of $f(\Phi)$ becomes higher. The predicted spectral tilt $n_s$ and tensor-to-scalar ratio $r$ are given by,
\begin{align}
n_s&\sim1-\frac{2n+1}{n+1}\frac{1}{N},\\
r&\sim\frac{8n}{n+1}\frac{1}{N},
\end{align}
where $N$ denotes the number of e-foldings and we have omitted smaller corrections. With a sufficiently large $n$, the scalar potential~(\ref{Vc}) asymptotes to $\phi^2$, which is favored by the BICEP2 data~\cite{Ade:2014xna}. This model can be tested by forthcoming experiments.
\subsection{Starobinsky inflation }\label{Starobinsky}
Next, we choose the following function $f(\Phi)$~\cite{Kallosh:2013lkr},
\begin{align}
f(\Phi)=\lambda(1-e^{-a\Phi}),
\end{align}
where $a$ and $\lambda$ are real constant parameters. Then, the canonical normalized complex scalar $\varphi$ is
\begin{align}
\varphi=4\sqrt{C}\lambda\left(\Phi+\frac{1}{a}e^{-a\Phi}\right)\sim4\sqrt{C}\lambda\Phi.
\end{align} 
Here, we identify $\sqrt{2}{\rm Re}~\varphi=\phi$ as the inflaton. Then, the effective potential is given by
\begin{align}
V\sim\lambda^2 (1-e^{-\frac{\tilde{a}}{\sqrt{2}}\phi})^2,
\end{align}
where $\tilde{a}=(4\sqrt{C} \lambda)^{-1}a$ and we only write down the leading terms. The resultant potential is same with one in the Starobinsky model~\cite{Kallosh:2013lkr}, and the predicted spectral tilt~$n_s$ and tensor-to-scalar ratio~$r$ are as follows,
\begin{align}
n_s&\sim 1-\frac{2}{N},\label{Stilt}\\
r&\sim \frac{16}{\tilde{a}^2N^2} \label{Sr},
\end{align} 
where $N$ denotes the number of e-foldings. The predicted $r$ is very small when $\tilde{a}\sim\mathcal{O}(1)$. For $N\sim 60$, the values of $(n_s,r)$ are compatible with the Planck2013 result~\cite{Ade:2013uln}, however, this model may be excluded if the result from BICEP2~\cite{Ade:2014xna} is confirmed by other experiments.

We also note that the mass of ${\rm Im}\varphi$ is much smaller than the Hubble scale during inflation although its minimum is located its origin. Then, the light direction can be a curvaton as in the case of the simplest chaotic inflation in SUGRA~\cite{Kawasaki:2000yn,Demozzi:2010aj}, if the decay of ${\rm Im}\varphi$ occurs after that of the inflaton~${\rm Re}\varphi$. Further investigation of such a case would be interesting but it is beyond the scope of this letter.
\section{Conclusion}\label{con}
We have proposed a new class of inflation models with a SUSY derivative term in SUGRA.  We have found that the kinetic terms of chiral multiplets are included in SUSY derivative terms, even if the multiplets are absent in K\"ahler potential terms. By virtue of the absence of the inflaton in $K$, the $\eta$ problem does not exist in our models. It is also remarkable that the action contains the higher order terms of $X=-\partial_\mu \phi \partial^\mu \phi/2$, however, their contributions are negligible and the effective action is given by the scalar system with the standard kinetic term and the scalar potential. 

We have shown two inflation models in our setup. In Sec.~\ref{chaotic}, we discussed the chaotic inflation model. It is surprising that the effective potential is restricted from the linear to the quadratic potential regardless of the superpotential containing the arbitrary power of $\Phi$. Therefore this model can be tested by forthcoming experiments.

On the other hand, we have also constructed the Starobinsky type inflation model in Sec.~\ref{Starobinsky}. In contrast to the chaotic type model, it predicts the very small value of the tensor-to-scalar ratio $r$ which is compatible with the Planck2013 result. As discussed in Sec.~\ref{Starobinsky}, this model contains the light scalar ${\rm Im}\varphi$ which may produce additional adiabatic perturbations. If it the case, the predicted cosmological parameters shown in Eq.(\ref{Stilt}) and (\ref{Sr}) can be changed. That is an interesting possibility but it requires more detailed investigation of other sectors. 

In this work, we have only discussed the dynamics of the scalar sector, however we need to take fermions into account to discuss the universe after inflation. That will be our future work. 
\section*{Acknowledgements}
The authors would like to thank Hiroyuki Abe for useful discussion and comments.
Y.Y. was supported by JSPS Research Fellowships for Young Scientists No. 26-4236 in Japan.
\appendix
\section{Cosmological parameters in our models}\label{A}
We show the cosmological parameters in our models here. In general, due to the existence of the higher order terms of $X=-\partial_\mu \phi \partial^\mu \phi/2$, the values of the spectral tilt and the tensor-to-scalar ratio are slightly different from the models without such higher order terms. However, the modification is negligible in the slow-roll inflation models as shown in the later.

Our models discussed in Sec.~\ref{model} are effectively single inflation models, and the action of the inflaton $\phi$ is given by
\begin{align}
\mathcal{L}=\frac{X}{A^2}-V(\phi)\left(\frac{2}{A}-\frac{1}{A^2}\right)\equiv P(X,\phi),
\end{align}
where $X=-\partial_\mu \phi \partial^\mu \phi/2$, $V(\phi)$ denotes the scalar potential, and $A=1+V^{-1}X$.

The Einstein equations and the equation of motion of $\phi$ in the flat Friedmann-Lema\^itre-Robertson-Walker background $g_{\mu\nu}={\rm diag}(-1,a(t), a(t),a(t))$ are given by
\begin{align}
3H^2&=2XP_X-P,\\
&\dot{H}=-XP_X,\\
(P_X+2XP_{XX})\ddot{\phi}+&3HP_X\dot{\phi}+2XP_X-P_\phi=0,
\end{align}
where the subscripts denote the derivative with respect to $X,\phi$, the dot denotes the time derivative, and $H=\frac{\dot{a}}{a}$ is the Hubble parameter. 

The scalar spectral tilt $n_s$ and the tensor-to scalar ratio $r$ are described by the slow-roll parameter $\epsilon$, $\eta$ and the variation of the speed of sound $s$ defined by
\begin{align}
\epsilon&\equiv -\frac{\dot{H}}{H^2},\\
\eta&\equiv\frac{\dot{\epsilon}}{H\epsilon},\\
s&\equiv\frac{\dot{c}_s}{Hc_s},
\end{align}
where $c_s^2\equiv P_X/(P_X+2XP_{XX})$ which is the speed of sound of the scalar perturbation. By using them, we can express $(n_s,r)$ as follows (see e.g.~\cite{Kobayashi:2011nu}),
\begin{align}
n_s&=1-2\epsilon -\eta-s,\\
r&=16\epsilon.
\end{align}

In slow-roll inflation cases, those parameters can be effectively written with the slow-parameter of the scalar potential $\epsilon_V$ and $\eta_V$ defined by
\begin{align}
\epsilon_V&\equiv\frac{1}{2}\left(\frac{V_\phi}{V}\right)^2,\\
\eta_V&\equiv\frac{V_{\phi\phi}}{V}.
\end{align}
We assume $|\epsilon_V|,|\eta_V|\ll1$, and then the set of parameters $(\epsilon, \eta,s)$ in our models are 
\begin{align}
\epsilon&\sim\epsilon_V(1+2\epsilon_V),\\
\eta&\sim-2\eta_V+4\epsilon_V -8\eta_V+\frac{17}{3}\epsilon_V^2,\\
s&\sim-\frac{2}{3}\eta_V\epsilon_V+\frac{4}{3}\epsilon_V^2,
\end{align}
where we omit the higher corrections $\mathcal{O}(\epsilon_V^3,\eta_V^3,\cdots)$. Therefore, the cosmological parameters $(n_s,r)$ are given by
\begin{align}
n_s&\sim 1+2\eta_V-6\epsilon_V+\frac{26}{3}\eta_V\epsilon_V-11\epsilon_V^2\nonumber\\
&\sim1+2\eta_V-6\epsilon_V,\\
r&\sim16\epsilon_V+32\epsilon_V^2\sim16\epsilon_V.
\end{align}
Those are same with ones in the standard slow-roll inflation models. 

\end{multicols}
\end{document}